\documentclass[utf8]{frontiersFPHY} % for Physics and Applied Mathematics and Statistics articles

\setcitestyle{square} % for Physics and Applied Mathematics and Statistics articles
\usepackage{url,hyperref,lineno,microtype,subcaption}
\usepackage[onehalfspacing]{setspace}
\usepackage{breakurl}
\usepackage{amsmath,amssymb,bm}
\usepackage{graphicx}
\usepackage{epsfig}
\usepackage{natbib}
\newcommand{\aap}{    {\it Astron. Astrophys.}}
\newcommand{\aaps}{   {\it Astron. Astrophys. Suppl.}}

\newcommand{\apj}{    {\it Astrophys. J.}}
\newcommand{\apjl}{   {\it Astrophys. J. Lett.}}

\newcommand{\apjs}{   {\it Astrophys. J. Suppl. Ser.}}

\newcommand{\jgr}{    {\it J. Geophys. Res.}}

\newcommand{\solphys}{{\it Solar Phys.}}

\newcommand{\ssr}{    {\it Space Sci. Rev.}}

\def\keyFont{\fontsize{8}{11}\helveticabold }
\def\firstAuthorLast{Song {et~al.}} %use et al only if is more than 1 author
\def\Authors{Hongqiang Song\,$^{1,2*}$, Qiang Hu\,$^{3}$, Xin Cheng\,$^{4}$, Jie Zhang\,$^{5}$, Leping Li\,$^{2}$, Ake Zhao\,$^{6}$, Bing Wang\,$^{1}$,Ruisheng Zheng\,$^{1}$, and Yao Chen\,$^{1}$}
\def\Address{$^{1}$Shandong Provincial Key Laboratory of Optical Astronomy and Solar-Terrestrial Environment, and Institute of Space Sciences, Shandong University, Weihai, China \\
$^{2}$CAS Key Laboratory of Solar Activity, National Astronomical Observatories, Chinese Academy of Sciences, Beijing, China \\
$^{3}$Department of Space Science and CSPAR, University of Alabama in Huntsville, Huntsville, Alabama, USA\\
$^{4}$School of Astronomy and Space Science, Nanjing University, Nanjing, China \\
$^{5}$Department of Physics and Astronomy, George Mason University, Fairfax, Virginia, USA\\
$^{6}$College of Physics and Electric Information, Luoyang Normal University, Luoyang, China}
\def\corrAuthor{Hongqiang Song}

\def\corrEmail{hqsong@sdu.edu.cn}

\begin{document}
\onecolumn
\firstpage{1}

\title[The Inhomogeneity of Composition]{The Inhomogeneity of Composition along the Magnetic Cloud Axis}

\author[\firstAuthorLast ]{\Authors} %This field will be automatically populated
\address{} %This field will be automatically populated
\correspondance{} %This field will be automatically populated

\extraAuth{}% If there are more than 1 corresponding author, comment this line and uncomment the next one.
%\extraAuth{corresponding Author2 \\ Laboratory X2, Institute X2, Department X2, Organization X2, Street X2, City X2 , State XX2 (only USA, Canada and Australia), Zip Code2, X2 Country X2, email2@uni2.edu}

\maketitle

\begin{abstract}

%%% Leave the Abstract empty if your article does not require one, please see the Summary Table for full details.
%\section{}
Coronal mass ejections (CMEs) are one of the most energetic explosions in the solar system. It is generally accepted that CMEs result from eruptions of magnetic flux ropes, which are dubbed as magnetic clouds in interplanetary space. The composition (including the ionic charge states and elemental abundances) is determined prior to and/or during CME eruptions in the solar atmosphere, and does not alter during magnetic cloud propagation to 1 AU and beyond. It has been known that the composition is not uniform within a cross section perpendicular to magnetic cloud axis, and the distribution of ionic charge states within a cross section provides us an important clue to investigate the formation and eruption processes of flux ropes due to the freeze-in effect. The flux rope is a three-dimensional magnetic structure intrinsically, and it remains unclear whether the composition is uniform along the flux rope axis as most magnetic clouds are only detected by one spacecraft. In this paper we report a magnetic cloud that was observed by ACE near 1 AU on 1998 March 4--6 and Ulysses near 5.4 AU on March 24--28 sequentially. At these times, both spacecraft were located around the ecliptic plane, and the latitudinal and longitudinal separations between them were $\sim$2.2$^{\circ}$ and $\sim$5.5$^{\circ}$, respectively. It provides us an excellent opportunity to explore the axial inhomogeneity of flux rope composition, as both spacecraft almost intersected the cloud center at different sites along its axis. Our study shows that the average values of ionic charge states exhibit significant difference along the axis for carbon, and the differences are relatively slight but still obvious for charge states of oxygen and iron, as well as the elemental abundances of iron and helium. Besides the means, the composition profiles within the cloud measured by both spacecraft also exhibit some discrepancies. We conclude that the inhomogeneity of composition exists along the cloud axis.

%Our study also suggest that the composition can vary along the flux-rope axis, and this should be considered in the three-dimensional magnetohydrodynamic simulations of flux-rope eruptions.

\tiny
 \keyFont{ \section{Keywords:} coronal mass ejection, magnetic flux rope, interplanetary coronal mass ejection, magnetic cloud, ionic charge state, elemental abundance} %All article types: you may provide up to 8 keywords; at least 5 are mandatory.
\end{abstract}

\twocolumn

%\citep{chenpengfei11,webb12,chengxin17,guoyang17}\citep{gosling91,xumengjiao19}\citep{cannon13,riley18}

\section{Introduction}
Coronal mass ejections (CMEs) are an energetic explosive phenomenon in the solar atmosphere \citep{chenpengfei11,webb12,chengxin17,guoyang17}, and they are called interplanetary coronal mass ejections (ICMEs) after leaving the corona. When ICMEs interact with the Earth's magnetosphere, they can cause geomagnetic storms \citep{gosling91,zhangjie07,xumengjiao19} and influence the normal work of high-tech equipments, such as the satellites, power grids and GPS navigation systems \citep{cannon13,riley18}. Therefore, it is of great significance to grasp the trigger mechanisms and eruption processes of CMEs.

The researchers of solar physics community have reached a consensus that CMEs result from eruptions of magnetic flux ropes (MFRs), which refer to a volumetric current channel with the helical magnetic field lines wrapping around the central axial field \citep[e.g.,][]{gibson06a,canou10}. In white-light coronagraph images, CMEs often exhibit a three-part structure, i.e., a bright front, a dark cavity and a bright core \citep{illing85}. The cavity and core have been considered as the MFR cross section and erupted filament, respectively, for several decades. However, recent studies clearly demonstrate that both the filaments and hot-channel MFRs can appear as the bright core \citep{howard17,song17b,song19a,song19b}. The hot channels are first revealed through extreme-ultraviolet passbands sensitive to high temperatures (e.g., 131 and 94 \AA) \citep{zhangjie12}, and they can also be observed in hard X-ray \citep{sahu20} and microwave \citep{wuzhao16} images. Researchers also suggest that the dark cavity corresponds to a low-density region with sheared magnetic field in the early eruption stage \citep{song19b}.

Both theoretical and observational studies reveal that MFRs can form prior to \citep{chen96,linjun00,zhangjie12,patsourakos13,chengxin13b} and during \citep{mikic94,antiochos99,song14a,ouyang15,wangwensi17} solar eruptions, while they might exist before eruptions in more events \citep{ouyang17}. The numerical simulations demonstrate that the repetitive magnetic reconnections could play an important role during the MFR evolution \citep{kumar16}. The remote-sensing observations have been widely used to investigate the MFR formation process \citep{song14a,chengxin15,zhengruisheng20}. The charge states within ICMEs are frozen-in near the Sun \citep{owocki83} and the relative abundances of elements with different first ionization potentials (FIPs) are different obviously in the corona and photosphere \citep{laming15,vadawale21}. As the composition does not alter during CME propagation to 1 AU and beyond \citep{song20b}, the in situ data are also employed to analyze the MFR formation \citep{song16,wangwensi17,huangjia18} and plasma origin \citep{song17a,fuhui20} of CMEs. So far the most complete composition data of ICMEs are provided by the solar wind ion composition spectrometer (SWICS) aboard the advanced composition explorer (ACE) and Ulysses, which can provide the charge states and elemental abundances of $\sim$10 elements \citep{lepri01}.

When an ICME has its nose pass through a spacecraft, the MFR will be detected as a magnetic cloud (MC) \citep{burlaga81,syed19,song20a}. This is schematically shown in Figure 1(a) (also see \citep{gopalswamy06,kim13} for a similar cartoon), where the purple arrow depicts a spacecraft trajectory crossing one ICME through its nose portion as marked with the blue rectangle. Figure 1(b) displays the MFR within the rectangle, and the green dots represent the center of each cross section. The black, blue and red arrows depict three different trajectories.

Several statistical studies have been conducted on ICME composition. Huang et al. \citep{huangjin20} analyzed the composition inside 124 MCs and reported that fast MCs have higher charge states and relative elemental abundances (except the C/O) than slow ones. Owens \cite{owens18} analyzed the charge states of carbon, oxygen, and iron within 215 ICMEs, including 97 MCs and 118 non-cloud events, and found that MCs exhibit higher ionic charge states than non-cloud events. Zurbuchen et al. \cite{zurbuchen16} performed a comprehensive analysis of the elemental abundances of 310 ICMEs from 1998 March to 2011 August. They reported that the abundances of low-FIP elements within ICMEs exhibit a systematic increase compared to the solar wind, and the ICMEs with elevated iron charge states possess higher FIP fractionation than the other ICMEs. Very recently, Song et al. \citep{song21a} reported that all the ICME compositions possess the solar cycle dependence.

In the meantime, some attentions are paid on the composition distribution inside each MC. Song et al. \citep{song16} found that the average values of iron charge states ($<$Q$_{Fe}$$>$) can present four regular profiles along the spacecraft trajectories throughout MCs, i.e., (i) a bimodal profile with both peaks higher than 12+, (ii) a unimodal profile with peak higher than 12+, (iii) and (iv) the $<$Q$_{Fe}$$>$ profile remains beyond and below 12+ throughout the spacecraft trajectory inside an MC, respectively. Their studies demonstrated that the charge states can be nonuniform within the cross section of a specified MC, and suggested that the above profiles are tightly correlated with both the impact factor of spacecraft trajectories and the formation process of MFRs. For example, the bimodal profile implies that the MFR exists prior to eruption, see Figure 8 in \citep{song16} for more details. In addition, the elemental abundances are not uniform within one cross section either \citep{song17a}. Therefore, a spacecraft can detect different composition profiles when it cross one MC along the blue and black arrows as shown in Figure 1(b), which are located in the same cross section perpendicular to the axis but with different impact factors. However, whether the inhomogeneity of composition exists along the MC axis remains unclear because most MCs are detected only by either ACE or Ulysses. Given the MC is a three-dimensional (3D) structure intrinsically, the axial distribution of composition can reveal whether different portions along the MFR axis experience different eruption processes in the corona.

In this paper, we report an intriguing event, in which an MC was observed by both ACE on 1998 March 4--6 and Ulysses on March 24--28. At these times, both spacecraft were located around the ecliptic plane, and the latitudinal and longitudinal separations between them were $\sim$2.2$^{\circ}$ and $\sim$5.5$^{\circ}$, respectively. The Grad-Shafranov (GS) reconstruction \citep{huqiang02,huqiang17a} demonstrated that the MC axis oriented in an approximate east-west direction with the axis direction at Ulysses being titled slightly away from that at ACE, and both spacecraft almost intersected the MC center \citep{dudan07}. This implies that the two spacecraft crosses the MC along two trajectories resembling the black and red arrows in Figure 1(b), respectively, and provides us an excellent opportunity to explore whether the composition is uniform along the axis. We introduce the data in Section 2, and give the observations in Section 3. Section 4 presents the conclusion and discussion.

\section{Data}
The data used in this paper are provided by several payloads on board the ACE and Ulysses spacecraft. ACE is in a halo orbit around the first Lagrangian point between the Earth and the Sun since it was launched in 1997. Ulysses was launched in 1990 and entered an elliptical and heliocentric orbit with an aphelion at $\sim$5.4 AU from the Sun and a perihelion distance of $\sim$1.34 AU. Magnetic field data are provided by the ACE/MAG \citep{smith98} and Ulysses/Magnetic field \citep{balogh92} instruments. The bulk solar wind properties and the helium abundances are from the Solar Wind Electron Proton Alpha Monitor (SWEPAM) \citep{mccomas98} on board ACE and the Solar Wind Observations Over the Poles of the Sun (SWOOPS) \citep{bame92} on board Ulysses. The SWICS instruments on board both spacecraft \citep{gloeckler98,gloeckler92} offer the composition of heavy ions.

\section{Observations}
The criteria used to identify MCs near 1 AU mainly include the enhanced magnetic field strength, smoothly changing of magnetic-field direction, declining profile of solar-wind velocity, low proton temperature (or low plasma $\beta$), and elevated He$^{2+}$/H$^{+}$ ratio \citep{burlaga81,cane03,chiyutian16}. ACE detected an MC on 1998 March 4--6 as shown in Figure 2. The purple vertical dashed line denotes the shock driven by the ICME, and the two purple dash-dotted lines demarcate the MC boundaries.

Figure 2(a) shows the total magnetic field strength and its three components in RTN coordinate, where the X-axis (R) points from Sun center to spacecraft, the Y-axis (T) is the cross product of solar rotational axis and X axis, lying in the solar equatorial plane towards the west limb, and the Z-axis (N) is the cross product of X and Y axes. The total magnetic field strength (black) increased obviously compared to the background solar wind and the Bn component (blue) changed its direction gradually within the MC, which are the typical features of MCs. Figures 2(b)--(d) present the velocity, density, and temperature of the ICME sequentially. The declining profile of velocity indicates that the MFR is expanding.

Ulysses detected an MC on March 24--28 \citep{dudan10} as shown in Figure 3, where the magnetic field, velocity, density, and temperature are presented from top to bottom panels sequentially. The velocity profile in Figure 3(b) shows that the MC keeps expansion during the propagation to 5.4 AU. Due to the continuous expansion, the total magnetic field intensity within this MC decreased obviously near 5.4 AU compared to $\sim$1 AU, see Figures 2(a) and 3(a). A shock exists within the MC as depicted with the red arrows in Figures 3(a) and 3(b), and the MC rear boundary can be identified through the He$^{2+}$/H$^{+}$ ratio and the plasma $\beta$ value \citep{dudan07}. Note that the shock does not influence our analyses about the ionic charge states and elemental abundances.

Previous studies \citep{skoug00,dudan07} have confirmed that the MC displayed in Figure (3) corresponds to that in Figure (2). Skoug et al. \citep{skoug00} fitted both MCs using a force-free model of magnetic field \citep{lepping90}, and found that their central speed and cloud axis direction were very similar. The increase in MC diameter between 1 and 5.4 AU was also consistent with an expanding MC. Besides, both MCs had left-handed field structure and contained the similar magnetic fluxes, which were further confirmed by Du et al. \citep{dudan07} with the GS reconstruction technique. In addition, Du et al. \citep{dudan07} input the plasma and magnetic field data observed by ACE to their magnetohydrodynamic model to simulate the MC propagation and evolution to the Ulysses location. They compared the model predictions and the Ulysses observations, and identified further that Ulysses and ACE observed the same MC. As mentioned, the ACE (at $\sim$1 AU) and Ulysses (at $\sim$5.4 AU) were located near the ecliptic plane with a latitudinal separation of $\sim$2.2$^{\circ}$ and a longitudinal separation of $\sim$5.5$^{\circ}$ when they detected the MC. The GS reconstruction showed that the MC axis oriented in an approximate east-west direction, and both spacecraft almost intersected the MC center \citep{dudan07}, which support that ACE and Ulysses crossed the MC at different sites along its axis and provide us an excellent opportunity to explore whether the axial composition is uniform.

We compare the composition measured by both spacecraft in Figure 4, where the black and red lines represent the results of ACE and Ulysses, respectively. Please note that we only plot the composition within the MC, i.e, the left/right boundary of each panel corresponds to the MC start/end time. The ionic charge states (C$^{6+}$/C$^{5+}$, O$^{7+}$/O$^{6+}$, and $<$Q$_{Fe}$$>$) and elemental abundances (Fe/O and He$^{2+}$/H$^{+}$) are presented in Figures 4(a)--(e). The average values within the MC are also shown in each panel. The blue horizontal dashed lines represent the corresponding means in the slow solar wind during solar maximum \citep{lepri13} for reference and comparison.

Our study shows that the average values of composition within an MC can possess significant differences along the axis. For example, the C$^{6+}$/C$^{5+}$ ratio measured by Ulysses (3.04) is 12 times higher than that by ACE (0.23). In the meantime, the differences could be relatively slight for some compositions. For example, the O$^{7+}$/O$^{6+}$ ratio measured by Ulysses (0.41) is higher than ACE (0.34) by $\sim$21\%. The means of $<$Q$_{Fe}$$>$ detected by both spacecraft are nearly identical ($\sim$10). As to the elemental abundance, the Fe/O ratio by ACE (0.17) is $\sim$42\% higher than that by Ulysses (0.12), and the He$^{2+}$/H$^{+}$ ratio of ACE (0.093) is higher than Ulysses (0.053) by $\sim$75\%.

Besides the average values, the composition profiles measured by both spacecraft also exhibit discrepancy. Figure 4(a) shows that the C$^{6+}$/C$^{5+}$ of Ulysses elevated at the MC center, while the ACE profile did not exhibit the central peak. The O$^{7+}$/O$^{6+}$ of Ulysses presented a multi-peak profile, while ACE did not detect obvious peaks as shown in Figure 4(b). The He$^{2+}$/H$^{+}$ of ACE elevated in the second half as displayed in Figure 4(e), different from the profile of Ulysses that did not have large variation along the whole path. These can rule out the possibility that the inhomogeneity of composition is induced by the erosion \citep{ruffenach12} completely during propagation from 1 AU to 5.4 AU. Moreover, the erosion effect should be small for this event as both MCs have the similar magnetic fluxes as mentioned. The profiles of $<$Q$_{Fe}$$>$ and Fe/O measured by both spacecraft also exhibit some different fluctuation characteristics as displayed in Figures 4(c) and 4(d). The above results prove that the composition is inhomogeneous along the MC axis.

\section{Conclusion and Discussion}
An MC was detected by ACE at $\sim$1 AU and Ulysses at $\sim$5.4 AU sequentially during March 1998, when both spacecraft were located around the ecliptic plane. The latitudinal and longitudinal separations between them were $\sim$2.2$^{\circ}$ and $\sim$5.5$^{\circ}$, respectively. The GS reconstruction \citep{dudan07} showed that the axis oriented in an approximate east-west direction, and both spacecraft almost intersected the MC center, which provided an excellent opportunity to explore whether the composition is uniform along the axis. We compared the ionic charge states of carbon, oxygen, and iron (C$^{6+}$/C$^{5+}$, O$^{7+}$/O$^{6+}$, and $<$Q$_{Fe}$$>$), as well as the elemental abundances of iron and helium (Fe/O and He$^{2+}$/H$^{+}$) along the two trajectories. The results showed that the average values of C$^{6+}$/C$^{5+}$ exhibit significant difference along the axis, while the differences are relatively slight but still obvious for O$^{7+}$/O$^{6+}$, $<$Q$_{Fe}$$>$, Fe/O, and He$^{2+}$/H$^{+}$. Besides the means, the composition profiles within the MC measured by both spacecraft also exhibit obvious discrepancies. We conclude that the inhomogeneity of composition exists along the MC axis.

The magnetic field within the MC measured by Ulysses did not exhibit the obvious changing of direction compared with the measurements of ACE, see Figures 2(a) and 3(a). This might indicate that Ulysses passed through the ICME along a path a little far from the MC center than ACE. Figure 4(a) showed that Ulysses detected high C$^{6+}$/C$^{5+}$ at its central portion, which should also be observed by ACE if the composition is uniform along the MC axis. However, the C$^{6+}$/C$^{5+}$ profile of ACE did not present the elevated center. Therefore, if assuming there were some uncertainties about the spacecraft path in the GS reconstruction, it will not change our conclusion about the axial inhomogeneity of MC composition.

The charge states of carbon, oxygen and iron are frozen-in sequentially in the corona, i.e., the frozen-in altitudes of carbon and iron are the lowest and highest, respectively, in these three elements. For example, the carbon are frozen-in below 1.5 solar radii \citep{chenyao03,landi12}, while the iron around 3--4 solar radii \citep{buergi86,boe18}. Therefore, the obvious differences of C$^{6+}$/C$^{5+}$ along the MC axis imply that the different portions of MFR along the axis experience eruption processes with different physical parameters (e.g., temperature, density, and velocity) in the low corona. The similar values of $<$Q$_{Fe}$$>$ indicate that the physical parameters along the axis approached in the high corona. These should be taken into account in 3D simulations of CMEs. The axial inhomogeneity of elemental abundances implies that the abundances are not uniform throughout the MC source region on the Sun.

Our study demonstrated that the axial composition is nonuniform inside an MC, while we can not conclude that this large inhomogeneity exists within each MC. More events are necessary to investigate the inhomogeneity of composition along the MC axis, which needs a CME being detected by several spacecraft sequentially or simultaneously at different locations. This becomes more realizable as Solar Orbiter was launched in 2020 \citep{muller13}. Besides, Chinese solar physicists are proposing several space missions \citep{ganweiqun19a} to explore the Sun and solar eruption further. The Lay a Finger on the Sun \citep{linjun19} will launch a spacecraft to explore the solar eruption near the Sun, thus it will provide more MC cases that are measured sequentially near the Sun and around 1 AU combined with other spacecraft. The Solar Ring \citep{wangyuming20} plans to deploy six spacecraft, grouped in three pairs, on a sub-AU orbit around the Sun. The two spacecraft in each group are separated by $\sim$30$^{\circ}$ and every two groups by $\sim$120$^{\circ}$, which can provide more cases that are measured simultaneously by two or more spacecraft around the ecliptic plane. All of these missions will facilitate the studies of solar eruptions and other related issues.

\section*{Conflict of Interest Statement}
%All financial, commercial or other relationships that might be perceived by the academic community as representing a potential conflict of interest must be disclosed. If no such relationship exists, authors will be asked to confirm the following statement:
The authors declare that the research was conducted in the absence of any commercial or financial relationships that could be construed as a potential conflict of interest.

\section*{Author Contributions}
HS led the analysis and drafted the manuscript. QH contributed to further understand the GS reconstruction results in \citep{dudan07}. XC, JZ, LL, and AZ provided suggestions to improve the research. BW helped to plot Figure 1. RZ and YC contributed to improve the manuscript.

\section*{Funding}
This work is supported by the CAS grants XDA-17040507 and the NSFC grants U2031109, 11790303 (11790300) and 12073042. Hongqiang Song is also supported by the open research program of the CAS Key Laboratory of Solar Activity KLSA202107.

\section*{Acknowledgments}
We thank the referees for their constructive comments and suggestions which helped to improve the original manuscript considerably. We acknowledge the use of data provided by ACE and Ulysses missions. Hongqiang Song thanks Dr. Ying D. Liu for the helpful discussions.

\section*{Data Availability Statement}
The plasma and magnetic field data of both ACE and Ulysses are publicly available at the \textit{ACE} science center (\url{http://www.srl.caltech.edu/ACE/ASC/level2/index.html}) and the Ulysses Final Archive (\url{http://ufa.esac.esa.int/ufa/}), respectively.

%\bibliographystyle{frontiersinSCNS_ENG_HUMS} % for Science, Engineering and Humanities and Social Sciences articles, for Humanities and Social Sciences articles please include page numbers in the in-text citations

%\bibliographystyle{frontiersinHLTH&FPHY} % for Health, Physics and Mathematics articles
%\bibliography{myref}

%%% Make sure to upload the bib file along with the tex file and PDF
%%% Please see the test.bib file for some examples of references

\onecolumn
\section*{Figure captions}

%%% Please be aware that for original research articles we only permit a combined number of 15 figures and tables, one figure with multiple subfigures will count as only one figure.
%%% Use this if adding the figures directly in the mansucript, if so, please remember to also upload the files when submitting your article
%%% There is no need for adding the file termination, as long as you indicate where the file is saved. In the examples below the files (logo1.eps and logos.eps) are in the Frontiers LaTeX folder
%%% If using *.tif files convert them to .jpg or .png
%%%  NB logo1.eps is required in the path in order to correctly compile front page header %%%

\begin{figure}[h!]
\begin{center}
\includegraphics[width=10cm]{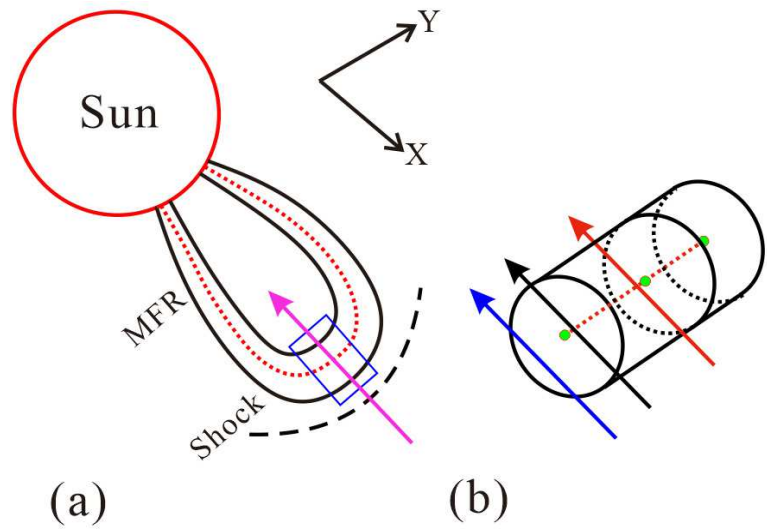}% This is a *.eps file
\end{center}
\caption{Schematic drawing of the spacecraft trajectory crossing an ICME. The black dashed and solid lines represent the shock and MFR, respectively. The red dotted lines delineate the MFR axis. The ICME nose portion is marked with the blue rectangle in Panel (a), which is enlarged for details in Panel (b). The blue, black, and red arrows describe the different trajectories of spacecraft, and the green dots denote the MC center.}\label{fig:1}
\end{figure}

\begin{figure}[h!]
\begin{center}
\includegraphics[width=10cm]{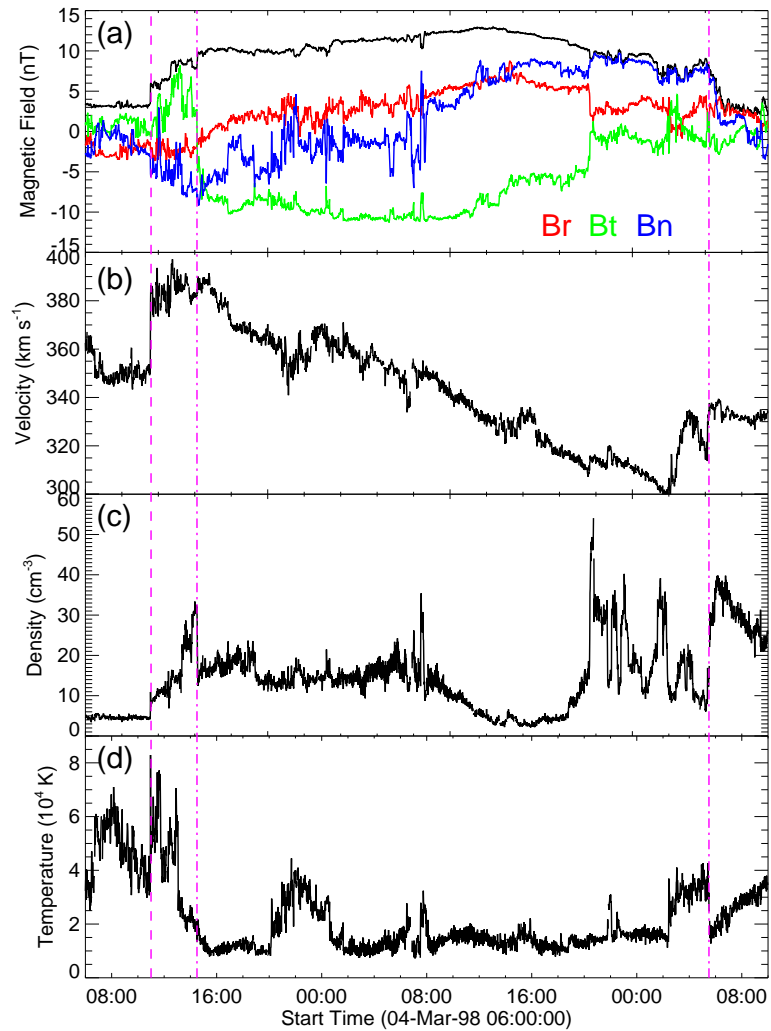}% This is a *.eps file
\end{center}
\caption{Magnetic field and solar wind parameters measured by ACE near 1 AU. (a) Total magnetic field strength (black) and its three components in RTN coordinate, (b)--(d) Velocity, density, and temperature of solar wind. The purple vertical dashed line denote the shock, and the dash-dotted lines demarcate the MC boundaries.}\label{fig:2}
\end{figure}

\begin{figure}[h!]
\begin{center}
\includegraphics[width=15cm]{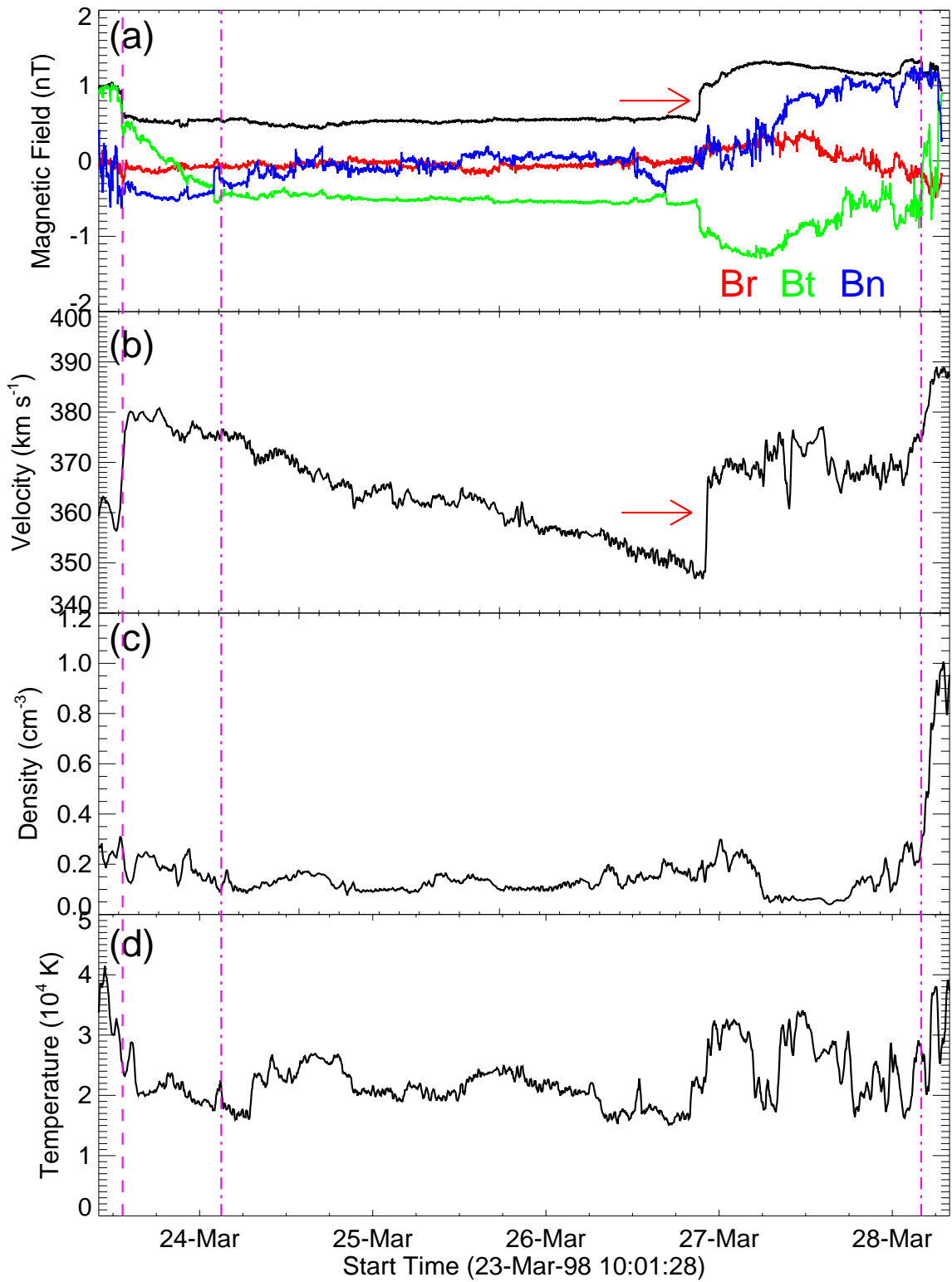}
\end{center}
\caption{Magnetic field and solar wind parameters measured by Ulysses near 5.4 AU. (a) Total magnetic field strength (black) and its three components in RTN coordinate, (b)--(d) Velocity, density, and temperature of solar wind. The purple vertical dashed line denote the shock, and the dash-dotted lines demarcate the MC boundaries. The red arrows in (a) and (b) depict the shock inside the MC.}\label{fig:3}
\end{figure}

\begin{figure}[h!]
\begin{center}
\includegraphics[width=15cm]{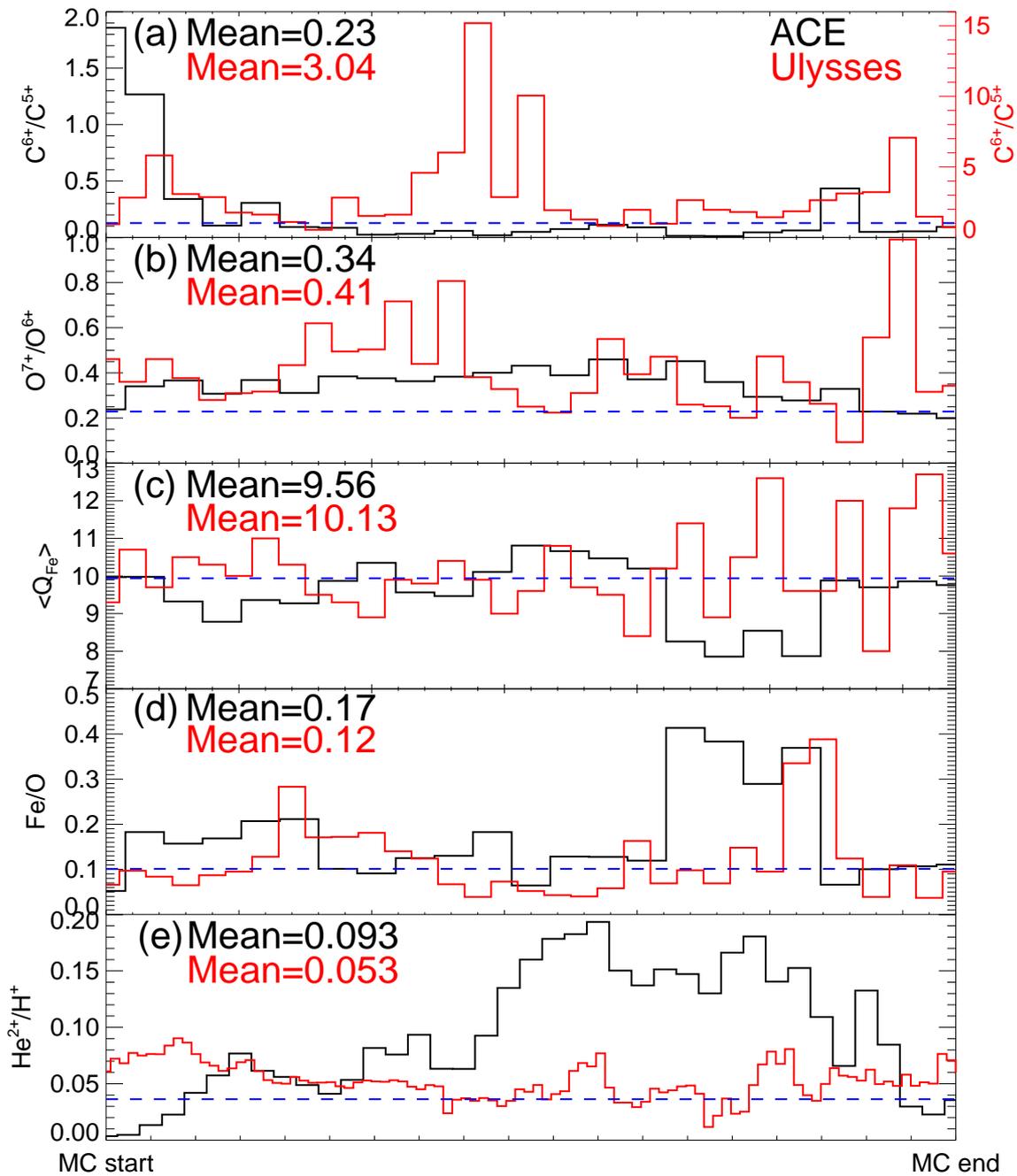}
\end{center}
\caption{Composition within the MC provided by SWICS aboard ACE (black) and Ulysses (red). Panels (a)--(e) show the C$^{6+}$/C$^{5+}$, O$^{7+}$/O$^{6+}$, $<$Q$_{Fe}$$>$, Fe/O, and He$^{2+}$/H$^{+}$ sequentially, and their average values are also presented in each panel. Note that the Ulysses values in Panel (a) correspond to the right ordinate. The blue horizonal dashed lines depict the corresponding means of slow wind during solar maximum \citep{lepri13}. The MC started from 14:30 UT on March 4 (3:00 UT on March 24) and ended at 5:30 UT on March 6 (4:00 UT on March 28) for ACE (Ulysses).}\label{fig:4}
\end{figure}

%%% If you are submitting a figure with subfigures please combine these into one image file with part labels integrated.
%%% If you don't add the figures in the LaTeX files, please upload them when submitting the article.
%%% Frontiers will add the figures at the end of the provisional pdf automatically
%%% The use of LaTeX coding to draw Diagrams/Figures/Structures should be avoided. They should be external callouts including graphics.

\end{document}